\documentclass[12pt]{article}
\usepackage[xdvi]{graphicx}
\usepackage{amssymb}
\newcommand{\bea}{\begin{eqnarray}}
\newcommand{\eea}{\end{eqnarray}}

\makeatletter
\@addtoreset{equation}{section}
\makeatother

\def\ignore#1{{}}

\begin{document}
\begin{titlepage}
\begin{flushright}
OU-HET 648/2009
\end{flushright}

\vspace{25ex}

\begin{center}
{\Large\bf 
Total-derivative supersymmetry breaking
}
\end{center}

\vspace{1ex}

\begin{center}
{\large
Naoyuki Haba and Nobuhiro Uekusa
}
\end{center}
\begin{center}
{\it Department of Physics, 
Osaka University \\
Toyonaka, Osaka 560-0043
Japan} \\
\end{center}


\vspace{3ex}

\begin{abstract}

On an interval compactification
 in supersymmetric theory, 
 boundary conditions for bulk fields 
 must be treated carefully. 
If they are taken arbitrarily following the requirement 
 that a theory is supersymmetric, 
 the conditions could give redundant constraints
 on the theory. 
We construct 
 a supersymmetric action integral on an interval 
 by introducing brane interactions 
 with which total derivative terms under the supersymmetry
 transformation become zero due to a cancellation.
The variational principle
 leads 
 equations of motion and also boundary conditions for 
 bulk fields, which 
 determine 
 boundary values of bulk fields. 
By estimating mass spectrum, 
  {\it spontaneous} supersymmetry breaking 
  in this simple setup can be realized 
  in a new framework. 
This supersymmetry breaking does not induce 
 massless R-axion, which 
 is favorable for phenomenology. 
It is worth noting that fermions in 
 hyper-multiplet, gauge bosons, and 
 the fifth-dimensional component of gauge bosons can 
 have zero-modes (while the other components are all massive 
 as Kaluza-Klein modes), 
 which fits 
 the gauge-Higgs unification scenarios.

\end{abstract}
\end{titlepage}

\section{Introduction}

Supersymmetry has been motivated
 from various viewpoints
 including 
 the gauge hierarchy problem
 and the gauge coupling unification as well as 
 string theory. 
Combining it with higher-dimensional theories also
 attract a lot of attention from 
 not only string theory but also phenomenologies.  
It is well known that a supersymmetric Lagrangian is
 invariant under the supersymmetry transformation up to 
 total derivative terms. 
In four dimensions,
 the action is supersymmetrically invariant 
 as fields fall to zero at spatial infinity. 
 
In higher-dimensional theories 
 with compactified extra-dimensions,
 such total derivative terms need to vanish
 for finite spaces with respect to extra-dimensions. 
Suppose that a total derivative term for a 
 five-dimensional example is 
 denoted as $\partial_y \Delta(x,y)$.
Integrating it out over $y$
 ($y_i\leq y \leq y_f$) 
 leads to $[\Delta(x,y_f)-\Delta(x,y_i)]$.
In a usual orbifold setup, 
 bulk fields have their parity eigenvalues associated with 
 orbifolding,  
 which derive 
 $\Delta(x,y_f)=\Delta(x,y_i)=0$, and 
 the action integral is supersymmetric. 
So, how about an interval compactification?
An interval compactification has been 
 intriguing for phenomenological 
 model building, such as Higgsless models~\cite{Csaki:2003zu}, 
 gauge-Higgs unification models~\cite{Agashe:2004rs}, 
 and so on,  
 since it has more varieties of boundary values 
 for bulk fields\footnote{
Also in a grand unified theory,  
 a five-dimensional interval compactification 
 can realize a direct reduction of 
 SO(10) to the standard model gauge group, 
 while an orbifold compactification 
 induces additional U(1)~\cite{Haba:2008hq}. 
}. 
In models on intervals, 
 it also seems to become supersymmetric if 
 bulk fields are assigned to have the same 
 boundary conditions as the orbifolding. 
In constructing the action integral on intervals,
 however,
 there are no inevitable conditions to fix boundary conditions
 for bulk fields unlike an orbifold compactification.
 If $y$-dependence of fields is specified beforehand
 so as to make an action integral supersymmetric, 
 it corresponds to a setup of a constrained system.
This constraint can be redundant in an interval 
 compactification, 
 while in an orbifold compactification
 it is the very orbifold condition that constrains
 the system. 
The assignment of boundary conditions for bulk fields 
 should be treated carefully.
The $y$-dependent profile for fields need to satisfy
 equations of motion and boundary conditions,  
 that are derived from 
 the variational principle
 for a supersymmetrically-invariant action integral. 
When 
 the values of
 $\Delta(x,y_f)$ and $\Delta(x,y_i)$ are arbitrary 
 in an interval compactification,
 it is nontrivial whether supersymmetry is 
 preserved or not. 
In order that total derivative terms vanish 
 in an interval compactification,
 another valid way may be to employ a cancellation.
Since 
 boundary contributions which cancel
 the variation $[\Delta(x,y_f)-\Delta(x,y_i)]$ 
 make the whole 
 theory supersymmetric, 
 the boundary action integral, 
 whose variation is $-[\Delta(x,y_f)-\Delta(x,y_i)]$, 
 would cancel the total derivative terms. 

In this paper,
 we construct 
 a supersymmetric action integral on an interval 
 by introducing brane interactions 
 with which total derivative terms under the supersymmetry
 transformation become zero due to a cancellation. 
The variational principle
 leads 
 equations of motion and also boundary conditions of 
 bulk fields, which 
 determine 
 boundary values of bulk fields. 
By estimating mass spectrum, 
 {\it spontaneous} supersymmetry breaking can be realized 
 in this simple setup. 
This supersymmetry breaking does not induce 
 massless R-axion, which 
 is favorable for phenomenology. 
It is worth noting that fermions in 
 hyper-multiplet, gauge bosons, and 
 the fifth dimensional component of gauge bosons can 
 have zero-modes (while the other components are all massive 
 as Kaluza-Klein (KK) modes), 
 which fits 
 the gauge-Higgs unification scenarios.

Our point is that
the action integral with boundary terms
is invariant under supersymmetry transformation
and that supersymmetry breaking
occurs without twist for boundary conditions.
Such a supersymmetry breaking without twist 
is supported by the results
in models given in Ref.~\cite{Diego:2005mu},
where the action integral is constructed with introduction of
the indices SU(2)${}_H$ as well as SU(2)${}_R$. 
Treatment of symmetric models with these indices has been
developed in Ref.~\cite{Diego:2006py}.
By comparison, we will construct a new supersymmetric
action integral
with boundary terms without introducing the index SU(2)${}_H$.
Our framework has 
three critical advantages.
Firstly, its consistency with earlier works is clear.
Our boundary terms are composed of the hyper-multiplet scalar field
and auxiliary field.
The form is expected from four-dimensional couplings
in Ref.~\cite{Mirabelli:1997aj}.
In the formulation with the index SU(2)${}_H$,
the fermion seems to have boundary terms~\cite{Diego:2005mu}.
To check consistency with the case without employing SU(2)${}_H$, 
several steps might be needed
\footnote{%
In Ref.~\cite{Diego:2005mu}, standard orbifold boundary conditions
are obtained as
$H^{\pm}(x,y_b +y) =\pm s_b t_b H^\pm (x, y_b -y)$
and $\psi_{L,R}(x,y_b +y)=\mp s_b \psi_{L,R} (x,y_b-y)$ with
$b=i,f$.
Here the twist parameters for SU(2)${}_H$ and SU(2)${}_R$
are denoted as $s_b$ and $t_b$, respectively.
When the SU(2)${}_H$ twist is trivial $s_b=\pm 1$,
fermions have zero mode and scalar masses depend on an SU(2)${}_R$ twist which
are consistent with our result.
}.
The second advantage is
simpleness of phenomenological application.
A hyper-multiplet in our framework is described by
an SU(2)${}_R$ complex scalar, a Dirac fermion and
an SU(2)${}_R$ complex auxiliary field.
Because our fermion is a usual Dirac fermion,
a simple introduction of quarks and leptons 
is possible directly in a realistic application.
Thirdly there is a clear indication 
for theoretical research.
In our supersymmetric formulation,
boundary terms for a hyper-multiplet are 
formed by only SU(2)${}_R$-charged fields.
This provides a transparent framework for developing
further research on relation between R symmetry and supersymmetry.
With these points and the new action integral, 
we can derive the observations also including the subjects
associated with R-axion and the gauge-Higgs unification 
scenarios.

The paper is organized as follows. 
In Sec.~\ref{sec:model}, 
 we construct a supersymmetric action integral 
 with a hyper-multiplet. 
In Sec.~\ref{solem}, equations of motion and 
 boundary conditions are derived 
 based on the variational 
 principle, and 
 solutions for the equations are represented. 
We show that it gives rise to 
  {\it spontaneous} supersymmetry breaking
  in a new framework.
We conclude in Sec.~\ref{sec:concl} with some remarks.

\section{Supersymmetric 
action \label{sec:model}}

In five dimensions, the number of 
 minimal supercharges is eight.
Associated with the supersymmetry,
 R-symmetry is SU(2)${}_R$ in the bulk. 
Simplectic Majorana fermions satisfy
 $\psi^i = \epsilon^{ij} C\bar{\psi}_j^T$,
 whose component fields are written
 as\footnote{
We follow the notation in Ref.~\cite{Hebecker:2001ke}.
}
\bea
  \psi^1 &\!\!\!=\!\!\!& \left(\begin{array}{c}
     (\psi_L)_\alpha \\
     (\bar{\psi}_R)^{\dot{\alpha}} \\
     \end{array}\right) , \qquad
  \psi^2 =\left(\begin{array}{c}
     (\psi_R)_{\alpha} \\
     -(\bar{\psi}_L)^{\dot{\alpha}} \\
     \end{array}\right)  .
\eea
Here $\epsilon^{12}=\epsilon_{21} =1$,
$C_{21}=-C_{12}=C_{34}=-C_{43}=1$
and $C_{AB}=0$ for the other matrix elements.
The gamma matrices are given by
\bea
   \gamma^M 
    =\left( \left(\begin{array}{cc}
       0 & \sigma^m \\
       \bar{\sigma}^m & 0 \\
       \end{array}\right) ,
       ~~
       \left(\begin{array}{cc}
       -i & 0 \\
       0 & i \\
       \end{array}\right)
       \right) ,
\eea
where $\{\gamma^M,\gamma^N\}=-2\eta^{MN}$.
Here  the capital letters $M$
run over $0,1,2,3,5$ and $\sigma^m =(1,\vec{\sigma})$,
$\bar{\sigma}^m =(1,-\vec{\sigma})$.
The five-dimensional spacetime is flat, and 
 the extra-dimensional coordinate is denoted also as $y$.
The fundamental region is taken as $0\leq y\leq L$.

For a hyper-multiplet, bulk fields are composed of
an SU(2)${}_R$ complex scalar $H^i$,
a Dirac fermion $\psi$ and 
an SU(2)${}_R$ complex auxiliary field $F_i$.
The kinetic Lagrangian for bulk fields are given by
\bea
   {\cal L} = -\partial_M H_i^\dag \cdot
     \partial^M H^i
     -{i\over 2}
   \left(\bar{\psi}\gamma^M \partial_M \psi
     -\partial_M \bar{\psi}\cdot \gamma^M  \psi\right)
     +F^{\dag i}F_i .
\eea
In order that the Lagrangian is invariant under supersymmetry
in the bulk,
supersymmetry transformation is given by
\bea
   \delta_\xi H^i &\!\!\!=\!\!\!&
   -\sqrt{2} \epsilon^{ij} \bar{\xi}_j \psi ,
\\
  \delta_\xi \psi
    &\!\!\!=\!\!\!&
    i\sqrt{2} \gamma^M \partial_M
     H^i \cdot \epsilon_{ij} \xi^j
     +\sqrt{2} F_i \xi^i ,
\\
 \delta_\xi F_i 
    &\!\!\!=\!\!\!&
      i\sqrt{2} \bar{\xi}_i \gamma^M \partial_M \psi .
      \label{lag1}
\eea
Here the transformation parameter $\xi^i$ 
is constant.
The conjugate transformation is given by
\bea
   \delta_\xi H_i^\dag &\!\!\!=\!\!\!&
   \sqrt{2} \epsilon_{ij}
    \bar{\psi} \xi^j ,
\\
 \delta_\xi \bar{\psi} 
   &\!\!\!=\!\!\!&
     i\sqrt{2} \partial_M H_i^\dag \cdot
      \epsilon^{ij} \bar{\xi}_j \gamma^M
      +\sqrt{2} F^{\dag i} \bar{\xi}_i ,
\\
  \delta_\xi F^{\dag i}
   &\!\!\!=\!\!\!&
    -i\sqrt{2} \partial_M\bar{\psi}\cdot
    \gamma^M \xi^i .
\eea
The supersymmetry transformation of 
the Lagrangian in Eq.(\ref{lag1}) is
\bea
  \delta_\xi {\cal L}
   &\!\!\!=\!\!\!&
  -\partial_M
   \left(
    \sqrt{2}\epsilon_{ij}\bar{\psi}
    \left(\eta^{MN} +{1\over 2}\gamma^M\gamma^N\right)
    \xi^j\partial_N H^i
    + {i\sqrt{2}\over 2}
     \bar{\psi}\gamma^M \xi^i F_i\right)
     +\textrm{H.c.} ,
\eea
which is decomposed in the four-dimensional derivative part 
and the extra-dimensional derivative part as
$\delta_\xi {\cal L}
= \partial_m \Delta^m +\partial_y \Delta_y$
with
\bea
  \Delta^m(x,y)
   &\!\!\!=\!\!\!&
  -\left(
    \sqrt{2}\epsilon_{ij}\bar{\psi}
    \left(\eta^{mn} +{1\over 2}\gamma^m\gamma^n\right)
    \xi^j\partial_n H^i 
    + {i\sqrt{2}\over 2}
     \bar{\psi}\gamma^m \xi^i F_i\right)
     +\textrm{H.c.} ,
\nonumber
\\
  \Delta_y(x,y) &\!\!\!=\!\!\!&
   -{1\over \sqrt{2}}
   \left(
    \bar{\psi}
    \left(
      \epsilon_{ij} \partial_y H^i  
    + i\gamma^5 F_j\right)  \xi^j
   + \epsilon_{ij}\bar{\psi}
    \left(  
 \gamma^5\gamma^m\partial_m H^i\right) \xi^j
    \right)
     +\textrm{H.c.} .
\label{2.11}
\eea
If extra-dimensional space were infinitely extended 
 and all the fields fell to zero at spatial infinity,
 the action integral is supersymmetrically invariant.
However, 
 it is quite non-trivial when 
 the extra-dimensional space is compactified in a 
 finite space, where 
 the action integral is transformed into
\bea
\int_0^L dy \  \delta_\xi {\cal L}
=\Delta_y(x,L)-\Delta_y(x,0).
\label{212}
\eea
Thus, it should be checked whether
 the total derivative terms with respect to $y$ 
 vanish in the finite space setups.

In an orbifold compactification, 
 taking $S^1/Z_2$ for example, 
 the spatial points of $y=0$ and $y=L$
 are fixed points with respect to the identification 
 $y\sim -y$ and $L+y \sim L-y$, respectively. 
Under the 
 orbifold parities $P_0$ and $P_1$
 at $y=0$ and $y=L$, 
 bulk fields are expected to have their parity eigenvalues
 in a usual setup. 
For example, let us take 
 orbifold parities as
 $P_0=P_1=+1$ for $(\xi_L^1, H^1, \psi_L, F_1)$
 and $P_0=P_1=-1$ for $(\xi_L^2, H^2 , \psi_R, F_2)$~\cite{Mirabelli:1997aj}.
Then, 
 from Eq.(\ref{2.11}), 
 the odd parity of 
 $(\xi_L^2, H^2 , \psi_R, F_2)|=0$ 
 and 
 the even parity of 
 $\partial_y H^1|=0$
 realize 
 $\Delta_y|=0$. 
So the action integral is 
 invariant under the supersymmetric transformation. 
Here 
 we do not say 
 orbifolding setup always has a supersymmetric action integral.  
In actual, a setup of Ref.\cite{vonGersdorff:2004cg} is 
 not the case, 
 where 
 twists at boundaries can make a situation complicated. 
Anyhow, we should remind again 
 that a boundary condition of $\partial_y H^1|=0$ 
 for a bulk field which has a finite boundary value 
 is needed to 
 make a supersymmetric action integral in the above case. 
Then, how about the situation in an interval (where 
 this condition is not 
 obvious)?

In an interval compactification,
 there are more boundary conditions which can be taken
 based on the variational principle~\cite{%
Csaki:2003dt,Sakai:2006qi}. 
Notice that 
 the $y$-dependence profiles of bulk fields must 
 be solutions 
 to equations of motion with 
 the boundary conditions.
Since the equations of motion are 
 derived from an action integral 
 with supersymmetry,
 it is natural that
 the action integral is constructed 
 without fixing boundary values of fields as initial conditions. 
Namely, 
 instead of fixing boundary values of fields,  
 we should construct a supersymmetric action integral at first, 
 and obtain bulk mode equations of motion, then 
 determine boundary values of bulk fields. 

For the supersymmetric action integral, 
 the simplest case is to take 
 a vanishing net variation for each boundary,
 which realizes  
 a supersymmetric action integral as shown in 
 Eq.~(\ref{212})\footnote{
Of course there is a more 
 complicated possibility of $\Delta_y(x,L)-\Delta_y(x,0)=0$ with 
 $\Delta_y(x,L)\neq 0$ and $\Delta_y(x,0)\neq 0$, 
 which will be done in a future work.  
}. 
Here we focus on a hyper-multiplet, while 
 a construction of a supersymmetric action integral
 of a vector multiplet was discussed 
 in Ref.~\cite{%
vonGersdorff:2004cg}.
Let us construct a supersymmetric action integral with 
 a vanishing net variation for each boundary
 in a general interval setup. 
{}From the dimensional counting,
candidates of Lagrangian terms on boundaries
for the cancellation are
\bea
  &&  \partial_y H_i^\dag \cdot H^i ,~~
    H_i^\dag \partial_y H^i ,~~
    \bar{\psi}\psi ,~~
    \bar{\psi}\gamma^5 \psi , ~~
    H^i F_i , ~~
    H_i^\dag F^{\dag i} ,
\nonumber
\\
  && \epsilon_{ij} (\partial_y H^i) H^j ,~~
  \epsilon^{ij} (\partial_y H_i^\dag) H_j^\dag ,~~
  \epsilon_{ij} H^i F^{\dag j} , ~~
  \epsilon^{ij} H_i^\dag F_j ,
\label{213}
\eea
and there are no other terms. 
We can show that 
 their supersymmetry transformations and
 $\Delta_y$ need
 a combination among Eq.~(\ref{213}) 
 in order to cancel between 
 supersymmetric transformations
 of bulk and boundary terms as 
\bea
   \Delta_y| 
     +\left[
      {1\over 2}A \delta_\xi (H_i^\dag 
      \partial_y H^i) 
      + {1\over 2}B\delta_\xi (\epsilon^{ij} H_i^\dag F_j) 
     +\textrm{H.c.} \right],
  \label{wa}
\eea
where $A$ and $B$ are numbers.
At $y=L$, Eq.~(\ref{wa}) is explicitly written as
\bea
 &&
   -{1\over \sqrt{2}} \bar{\psi}
     \left((1-A) \epsilon_{ij}
      \partial_y H^i 
      +(i\gamma^5 + B) F_j \right) \xi^j 
\nonumber
\\
  &&
    +{1\over \sqrt{2}}
      \epsilon_{ij} \partial_m \bar{\psi}\cdot
        (\gamma^5 +iB) \gamma^m H^i \xi^j
   +{1\over \sqrt{2}}
   \epsilon_{ij}  \partial_y \bar{\psi}\cdot
      (A+iB\gamma^5) H^i \xi^j .
      \label{eqsu}
\eea
Note that brane interactions are only for 
 scalar and auxiliary (SU(2)$_R$ non-singlet) fields, and there are
 no interactions for fermion (SU(2)$_R$ singlet) fields. 
At the other boundary $y=0$,
 similar terms are found in a parallel way.
In order to vanish Eq.~(\ref{eqsu}), 
 $A$ must satisfy 
 $A=1$, and $B$ must satisfy
\bea
    (i\gamma^5 + B) \xi^j =0 .
\eea
Since the value of $\gamma^5$ depends on
 the chirality of $\xi^j$ and $B$ is just a number,
 there is no solution of $B$ for general $\xi^j$.
Therefore,
  a whole four-dimensional
 ${\mathcal N}=2$ supersymmetry can not be preserved
 on boundaries,
 which is a well-known result as an origin of 
 a four-dimensional chiral theory from 
 a five-dimensional {\it vector-like} theory.  
Taking an eigenvalue of $\gamma^5$ for 
 $\xi^i$, $B$ has a solution, and then  
 a four-dimensional ${\mathcal N}=1$ supersymmetric theory 
 is obtained. 
Namely, 
if $\xi_L^i=0$ and $\xi_R^i\neq 0$ at $y=L$, 
 $-i\gamma^5 \xi^j = \xi^j$ 
 (for the notation $-i\gamma^5 \xi_R^i =\xi_R^i$), 
 which means that $B$ must satisfy $(-1+B)=0$, and 
 ${\mathcal N}=1$ supersymmetry is preserved 
 with $B=1$. 
If $\xi_R^i=0$ and $\xi_L^i\neq 0$ at $y=L$, 
 $-i\gamma^5 \xi^j = -\xi^j$, which means 
 $(1+B)=0$ and the 
 $B=-1$ solution preserves ${\mathcal N}=1$ supersymmetry. 
Hence,
 the supersymmetric action is given as
\bea
  S&\!\!\!=\!\!\!& \int d^4x dy
  \left( 
   -\partial_M H_i^\dag \cdot \partial^M H^i
   -{i\over 2} 
   \left(\bar{\psi}\gamma^M \partial_M \psi
     - \partial_M\bar{\psi}\cdot \gamma^M  \psi
     \right)
   +F^{\dag i} F_i 
   \right.
\nonumber
\\
   &&\left.
   + \left[
  {1\over 2} H_i^\dag
      (\partial_y H^i
       +B_1 \epsilon^{ij} F_j)\delta(y-L)
    -{1\over 2} H_i^\dag
      (\partial_y H^i + B_0 \epsilon^{ij}F_j)
        \delta(y) 
  +\textrm{H.c.} \right]\right) .
   \label{sact}
\eea
Here $B_s$ $(s=0,1)$ should satisfy
 $(i\gamma^5 + B_s)\xi^j =0$, which means 
 ${\mathcal N}=1$ supersymmetry is preserved 
 as $\xi_R^i\neq 0$ and $\xi_L^i= 0$ ($B_s=1$) or 
 $\xi_L^i\neq 0$ and $\xi_R^i= 0$ ($B_s=-1$).
It is worth noting that 
 the added boundary action
 in the action integral Eq.(\ref{sact})  
 is only formed by 
 the fields with charges of SU(2)${}_R$.

\section{Solving equations of motion
\label{solem}}

We have found the bulk and boundary action integrals
 invariant under supersymmetry
 in the previous section.  
Since we do not know the boundary values 
 for bulk fields on intervals, 
 it is natural to 
 construct a supersymmetric action integral at first, 
 and determine boundary values of bulk fields by 
 the variational principle 
 instead of fixing the values as the 
 initial condition\footnote{
Fixing boundary values as the initial condition 
 should correspond to fixing boundary conditions 
 of the model on interval. 
}. 
This is our standing point in this paper. 
Let us introduce 
 solutions for equations of motion and 
 boundary conditions from 
 the variational principle,
 and estimate boundary values of bulk fields.  

In the action integral Eq.(\ref{sact}),
 the direction of supersymmetry transformation can differ
 at each boundary depending on the values of $B_s$.
A possibility to keep global ${\mathcal N}=1$ supersymmetry
 would be to take the same direction $B_0=B_1$. 
For example, we choose $\xi_L^i\neq 0$ and $\xi_R^i= 0$ 
 at $y=0,L$ where $B_0=B_1=-1$. 
Then the action integral Eq.(\ref{sact}) becomes 
\bea
  S&\!\!\!=\!\!\!& \int d^4x dy
  \left( 
   -\partial_M H_i^\dag \cdot \partial^M H^i
   -{i\over 2} 
   \left(\bar{\psi}\gamma^M \partial_M \psi
     - \partial_M\bar{\psi}\cdot \gamma^M  \psi
     \right)
   +F^{\dag i} F_i 
   \right.
\nonumber
\\
   &&\left.
   +\partial_y \left(
  {1\over 2} H_i^\dag
      (\partial_y H^i
       - \epsilon^{ij} F_j)
  +\textrm{H.c.} \right)\right) ,
   \label{sact2}
\eea
where the boundary action is denoted as 
 a total derivative term. 
The Dirac fermion in hyper-multiplet  
 is SU(2)${}_R$ singlet and does not have 
 boundary interactions  
 as shown in Eq.(\ref{sact2}),  
 so that
 the existence of zero-mode 
 in the above boundary conditions
 is obvious. 
On the other hand, 
 for the SU(2)${}_R$ non-singlet fields, 
 whether they have zero-modes or not is 
 nontrivial since they have their boundary action integral.
Under the variations of SU(2)${}_R$ non-singlet fields as 
 $H^i \to H^i +\delta H^i$ and
 $F_i \to F_i +\delta F_i$,
 the variation of the action integral is given by
\bea
 \delta S 
  &\!\!\!=\!\!\!&
  {1\over 2}  \int d^4 x dy
    \left( 
     2\delta H_i^\dag \partial_M \partial^M H^i 
    +2\partial^M \partial_M H_i^\dag \cdot
     \delta H^i
 +2\delta F^{\dag i} \cdot F_i
   +2F^{\dag i}\delta F_i 
   \right.
\nonumber
\\
  && +\left[
     \delta H_i^\dag
       (-\epsilon^{ij} F_j)
        + H_i^\dag (-\epsilon^{ij} \delta F_j) \right]
        \delta (y-L) 
\nonumber
\\
  && +\partial_y (H_i^\dag \delta H^i
   \delta(y-L))
   -2\partial_y H_i^\dag \cdot \delta H^i \delta(y-L) 
   -H_i^\dag \delta H^i \partial_y \delta(y-L) 
\nonumber
\\
  && +  \partial_y (\delta H_i^\dag
       \cdot H^i \delta(y-L))
    -\delta H_i^\dag \cdot 
         \partial_y H^i \cdot \delta(y-L)
       -  \delta H_i^\dag
       \cdot H^i \partial_y \delta(y-L)
\nonumber
\\
  && + (-\epsilon_{ij} \delta F^{\dag j}) H^i \delta(y-L)
  +(\partial_y H_i^\dag
    - \epsilon_{ij} F^{\dag j}) \delta H^i \delta(y-L)
\nonumber
\\
   && \left.
     + (\delta(y) ~\textrm{terms}) \right) .
\eea
From this equation,
$\delta H_i^\dag$ terms mean
\bea
  &&   2\partial_M \partial^M H^i
       -\epsilon^{ij} F_j
        \delta (y-L) 
    +\epsilon^{ij} F_j
        \delta (y) 
\nonumber
\\
   &&    +H^i \delta(0) \delta(y-L)
    -\partial_y H^i \cdot \delta(y-L)
    -H^i \partial_y \delta(y-L)
\nonumber
\\
   &&    +H^i \delta(0) \delta(y)
    +\partial_y H^i \cdot \delta(y)
    +H^i \partial_y \delta(y)
   =0 ,
\eea
and 
$\delta F^{\dag i}$ terms mean
$2 F_i +\epsilon_{ij}H^j \delta(y-L)
-\epsilon_{ij}H^j\delta(y) =0$.
Combining these equations induces 
 three boundary conditions, 
\bea
   && 
   2F_i +\epsilon_{ij} H^i\delta(y-L) -\epsilon_{ij}
     H^j \delta(y) =0 ,
     \label{bc1}
\\
  &&\left. \left[ 4 \partial_y H^i
    +{3\over 2} H^i \delta(0)\right]\right|_{y=0} = 0,
   \qquad
   \left. \left[ -4 \partial_y H^i +{3\over 2} H^i \delta(0)
     \right]\right|_{y=L}
   =0 ,
     \label{yL}
\eea
and a bulk equation of motion, 
\bea
\partial_M \partial^M H^i=0. 
\label{36}
\eea
Here the behavior of fields near boundaries
 are treated as $\partial_y H^i|_{y=0} =
\lim_{\epsilon\to 0}\partial_y H^i|_{y=\epsilon}$ and
$\partial_y H^i|_{y=L}=
\lim_{\epsilon\to 0}\partial_y H^i|_{y=L-\epsilon}$.
With the mode expansion of $H^i(x,y)=\sum_n \phi^i_n(x) H^i_n(y)$,
 the bulk mode equation means
 $\partial_y^2 H_n^i =-m_n^2 H_n^i$.

Now we solve the equation of motion Eq.~(\ref{36}) under the
boundary conditions Eqs.~(\ref{bc1}) and (\ref{yL}).
From the bulk equation, a general solution apart from boundaries
is 
\bea
   H_n^i(y) = \sin (m_n y+\alpha) ,
    \label{gens}
\eea
up to the normalization.
Here $m_n$ and $\alpha$ are constants determined by 
 boundary conditions in the following.
Dependences on $i=1,2$ are omitted when no confusion arises. 
At $y=0$, substituting Eq.~(\ref{gens})
 into the first equation in Eq.~(\ref{yL}) gives
\bea
  4 m_n\cos \alpha +
   {3\over 2} \sin \alpha \cdot \delta(0) = 0 .
  \label{eq1}
\eea
If $\sin\alpha =0$, this equation means
 $m_n=0$, so that trivially $H^i(x,y)=0$.
Non-vanishing $H^i$ requires
 $\sin \alpha \neq 0$, which means 
 $\delta(0) = -(8/3) \cot \alpha \cdot m_n$.
At $y=L$, the second equation in Eq.~(\ref{yL}) with 
the above equation leads to
\bea
  {\tan m_nL \over m_nL} \simeq {16\over 3}
  {1\over L\delta(0)} .    
  \label{ml}
\eea
Since $\delta(0) \gg 1/L$, the mass eigenvalue is obtained as
\bea
  m_n = {n\pi\over L}\left(1+{16 \over
     3L \delta(0)}\right) 
  \simeq {n\pi \over L} .
\eea
Thus we find the solution
\bea
  H^i (x,y) =
   \sum_{n=1}^\infty
  \phi^{i}_n (x) \sin \left(
   m_n\left[y-{8\over 3\delta(0)}\right]\right) 
  \simeq \sum_{n=1}^\infty
   \phi_n^{i}(x) \sin (m_ny) ,
\eea
which means 
 the field $H^i(x,y)$ does not have zero-mode as 
 $m_n\neq 0$.
The absence of zero-mode is seen directly from 
 $\delta(0)$-term in Eq.~(\ref{yL}) and the bulk equation. 
The zero-mode of the bulk equation means a constant solution, 
 and 
 a constant $H^i$ does not fulfill Eq.~(\ref{yL})
 due to nonzero $\delta(0)$-terms which 
 correspond to 
 brane localized interactions of making supersymmetric action.
The mass splitting between $\psi$ and $H^i$ 
 is $\pi/L$, which 
 is characterized by the dimensional
 quantity $L$. 
(Here, $L$ is only one dimensionful parameter in 
 this model.)
 
We emphasize that
 the equations of motion with boundary conditions 
 remove 
 zero-mode for hyper-multiplet scalars.
This means that the action integral is supersymmetric but 
 the vacuum is not, that is, supersymmetry is 
 {\it spontaneously} broken. 
This mechanism of supersymmetry breaking 
 has been realized in a very simple formulation.
Relations between R-symmetry and supersymmetry breaking
 have been generally discussed~\cite{Nelson:1993nf}.
Usually, in four dimensions, 
 R-symmetry is needed for the spontaneous
 supersymmetry breaking, but there appears massless R-axion 
 which induces phenomenological difficulties.  
How about our setup in five dimensions?
In our
 present context 
 with R-symmetry which corresponds to 
 ${\mathcal N}=1$ supersymmetry of $B_0=B_1$, 
 this supersymmetry is {\it spontaneously} broken.  
Here massless R-axion is absent, since 
 there is no scalar source of R-axion in our setup. 
It does not contradict 
 the potential arguments of spontaneous
 supersymmetry breaking in four dimensions~\cite{Nelson:1993nf}.
Notice also that 
 gauginos have Dirac-type KK mass terms 
 although there is 
 R-symmetry. 
It is because KK mass is not a (simplectic)
 Majorana mass, so that this situation do not contradict 
 with the arguments in Ref.\cite{Nelson:1993nf}. 
Thus, gauginos can be massive without massless R-axion 
 in this setup.  
This mechanism sheds light on to 
 phenomenology. 
In a short summary, 
 the boundary interactions, which are introduced to 
 construct the supersymmetric action integral 
 in a five-dimensional theory, 
 realize the total derivative terms in the Lagrangian, 
 and ${\mathcal N}=2$ supersymmetry is completely 
 broken to ${\mathcal N}=0$ through the 
 equations of motion.   

In vector multiplets, gauginos and auxiliary fields
 are SU(2)${}_R$ non-singlet fields.
Gauginos, auxiliary fields, and 
 real scalar which is not the extra-dimensional
 component of gauge bosons, 
 have boundary terms for the supersymmetric action. 
Then, zero-modes exist only in  
 gauge bosons  
 and the fifth dimensional component of gauge bosons, if exist. 
This situation can be read from 
 Ref.~\cite{vonGersdorff:2004cg}, 
 where Scherk-Schwarz twist\cite{Scherk:1978ta}
 and boundary actions are taken into 
 account.
It can be shown that 
 a nonzero twist with a specific boundary condition 
 (not taking the same values of $B_0$ and $B_1$ as above)
 under a certain setup (Scherk-Schwarz twist), 
 SU(2)${}_R$ non-singlet fields can also have zero-mode. 
This means 
 ${\mathcal N}=1$ supersymmetry is preserved 
 after the compactification. 
Although our setup 
 induces supersymmetry breaking, 
 it is consistent with 
 Ref.\cite{vonGersdorff:2004cg}. 
It is because we do not take the 
 twists 
 at the boundary conditions. 
As for the vacuum energy, our 
 setup ({\it spontaneous} supersymmetry breaking vacuum) might have  
 higher magnitude than the setup of Ref.\cite{vonGersdorff:2004cg} 
 (supersymmetry preserving vacuum).  
However, 
 a dynamical mechanism of compactification 
 is still a mystery so that both setups are worth 
 analyzing carefully. 
Anyhow, 
 our setup is simple and 
 supersymmery breaking occurs without such a source
 as Scherk-Schwarz twist. 
And, 
 hyper-multiplets and vector multiplets
 have zero-mode for hyper-multiplet fermions and 
 five-dimensional gauge bosons. 
Since scalars of hyper-multiplets become heavy (nonzero-mode),
 the standard model Higgs field can not 
 be regarded as a field in hyper-multiplets. 
However, things go well if 
 the Higgs field can be
 identified as a part of the extra-dimensional component of 
 five-dimensional gauge bosons, because 
 it has zero-modes as mentioned above. 
This is a so-called gauge-Higgs unification model,
 and 
 the gauge hierarchy problem for quadratic divergence
 for Higgs boson mass is
 solved not by supersymmetry but by the gauge-Higgs unification.
Therefore, our {\it spontaneous} supersymmetry breaking
 gives a compelling way to extract just viable fields 
 in the non-supersymmetric gauge-Higgs 
 unification from a supersymmetric setup.

\section{Summary and discussions \label{sec:concl}}

We have constructed a five-dimensional supersymmetric action
 integral
 where total derivative terms play a role of canceling 
 the supersymmetry transformation of a boundary action integral.
After deriving equations of motion and boundary conditions,
 we have found solutions for these equations.
The solutions remove 
 zero-mode for hyper-multiplet scalars.
This means that the action is supersymmetric but 
 the vacuum is not, that is, supersymmetry is 
 {\it spontaneously} broken. 
This mechanism of supersymmetry breaking
 has been realized in a new formulation.
 
We have shown 
 that zero-mode exists only for fermions 
 in hyper-multiplets.
Combining our result with the case of vector multiplets
 in Ref.~\cite{vonGersdorff:2004cg}, 
 hyper-multiplets in the gauge theory 
 have zero-modes only in fermions of the hyper-multiplets and
 bosons (four-dimensional
 gauge bosons and scalars corresponding to the 
 fifth-dimensional component of gauge bosons). 
The mass of the Higgs boson depends
 on the compactification scale and a Wilson-line phase. 

For a model building, it would be also possible to change
 the setup to produce a smaller mass splitting than
 that of the order of $1/L$. 
This may occur if additional twist contributions 
 (suitable twists between $B_0$ and $B_1$, and 
 Scherk-Schwarz) are included.
In other words, supersymmetry may be broken at lower scales.
A mixing of supersymmetry breaking 
 from boundary action integrals and Scherk-Schwarz twists 
 can 
 reduce a magnitude of mass splitting. 
It has been shown that such a mixing can relax
 the lower bound to the lightest Higgs boson mass
 in a supersymmetric orbifold model~\cite{Uekusa:2008bd},
 even if $1/L$ is of the orders of magnitude larger than
 ${\cal O}(1)$~TeV.


\subsubsection*{Acknowledgments}

The authors thank Yoshio Koide, Nobuchika Okada, 
 and Koichi Yoshioka 
 for valuable conversations.
Especially, NH is grateful to Yoshiharu Kawamura for
 inspiring discussions in the early stage of this research.
This work is supported by Scientific Grants 
 from the Ministry of Education
 and Science, Grant No.~20244028, 
 No.~20540272, No.~20039006, and No.~20025004.


\begin{appendix}
  
\end{appendix}




\end{document}